\RequirePackage{fix-cm}
\documentclass[twocolumn,epjc3]{svjour3}
\RequirePackage{graphicx,subfigure}
\RequirePackage{multirow}
\RequirePackage{lineno}

\begin{document}
\title{Rejection of randomly coinciding events in Li$_2$$^{100}$MoO$_4$ scintillating bolometers using light detectors based on the Neganov-Luke effect} 
\titlerunning{RCD-Luke}

\author{D.M.~Chernyak\thanksref{addr1,e1}\and 
F.A.~Danevich\thanksref{addr1}\and 
L.~Dumoulin\thanksref{addr2}\and 
A.~Giuliani\thanksref{addr2,addr3,e2}\and
M.~Mancuso\thanksref{addr2,addr3,e3}\and 
P.~de~Marcillac\thanksref{addr2}\and
S.~Marnieros\thanksref{addr2}\and
C.~Nones\thanksref{addr4}\and
E.~Olivieri\thanksref{addr2}\and
D.V.~Poda\thanksref{addr1,addr2}\and
V.I.~Tretyak\thanksref{addr1,addr5}}
\thankstext{e1}{Presently at Kavli Institute for the Physics and Mathematics of the Universe (WPI), 
The University of Tokyo Institutes for Advanced Study, The University of Tokyo, 
Kashiwa, Chiba 277-8583, Japan}
\thankstext{e2}{e-mail: andrea.giuliani@csnsm.in2p3.fr}
\thankstext{e3}{Presently at Max-Planck-Institut f\"ur Physik, 80805 Munich, Germany}
\institute{Institute for Nuclear Research, MSP 03680 Kyiv, Ukraine \label{addr1}
\and
CSNSM, Univ. Paris-Sud, CNRS/IN2P3, Universit\'e Paris-Saclay, 91405 Orsay, France \label{addr2}
\and
DISAT, Universit\`a dell'Insubria, 22100 Como, Italy \label{addr3}
\and
CEA Saclay, DSM/IRFU, 91191 Gif-sur-Yvette Cedex, France \label{addr4}
\and
INFN, sezione di Roma, I-00185 Rome, Italy \label{addr5}}
\date{Received: date  / Accepted: date}
\maketitle
\begin{abstract}
Random coincidences of nuclear events can be one of the main background 
sources in low-temperature calorimetric experiments looking for neutrinoless 
double-beta decay, especially in those searches based on scintillating bolometers 
embedding the promising double-beta candidate $^{100}$Mo, because of the relatively 
short half-life of the two-neutrino double-beta decay of this nucleus. We show in this 
work that randomly coinciding events of the two-neutrino double decay of $^{100}$Mo 
in enriched Li$_2$$^{100}$MoO$_4$ detectors can be effectively discriminated by pulse-shape 
analysis in the light channel if the scintillating bolometer is provided with a Neganov-Luke 
light detector, which can improve the signal-to-noise ratio by a large factor, assumed 
here at the level of $\sim 750$ on the basis of preliminary experimental results obtained 
with these devices. The achieved pile-up rejection efficiency results in a very low contribution, of the order of $\sim 6\times10^{-5}$ counts/(keV$\cdot$kg$\cdot$y), to the background 
counting rate in the region of interest for a large volume ($\sim 90$~cm$^3$) Li$_2$$^{100}$MoO$_4$ detector. 
This background level is very encouraging in view of a possible use of the Li$_2$$^{100}$MoO$_4$ solution for a bolometric tonne-scale next-generation experiment as that proposed in the CUPID project.
\end{abstract}


\section{Introduction}

The double-beta ($2\beta$) decay is an extremely rare nuclear 
transition in those even-even nuclides where ordinary beta decay 
is either forbidden by conservation of energy or highly 
suppressed by a large spin change. While the two-neutrino mode of 
the decay ($2\nu2\beta$), being allowed in the Standard Model of 
particles (SM), was observed experimentally after long-time 
efforts (see, e.g., reviews \cite{Saakyan:2013,Barabash:2015}), 
the neutrinoless double-beta ($0\nu2\beta$) decay violates 
lepton number conservation \cite{Vergados:2012,Barea:2012,Rodejohann:2012} and is therefore forbidden in the framework of the SM. The process is 
considered as a unique way to investigate properties of neutrino 
and test many other hypothetical effects beyond the SM implying 
lepton number non conservation. 
The study of $0\nu2\beta$ decay can establish the 
Majorana nature of neutrino, help determine the 
scale of the neutrino mass, the neutrino-mass hierarchy and the 
Majorana CP-violating phases, check possible contribution of 
hypothetical right-handed currents admixture to weak interaction, 
the existence of Nambu-Goldstone bosons (majorons), and many other 
new-physics effects \cite{Deppisch:2012,Bilenky:2015,Pas:2015}. 

Despite almost seventy years of experimental activity, the 
$0\nu2\beta$ decay has not been observed yet. The most sensitive 
experiments give half-life limits on the level of 
$10^{24}-10^{26}$ y (see reviews 
\cite{Elliott:2012,Giuliani:2012a,Cremonesi:2014,Sarazin:2015,Delloro:2016}, 
and the recent KamLAND-Zen results~\cite{Kamland:2016}), 
which correspond to effective Majorana neutrino mass limits on the 
level of $\langle m_{\nu}\rangle \sim 0.1-1$ eV. The next-generation experiments  should explore the inverted region of the 
neutrino mass ($\langle m_{\nu}\rangle \sim 0.02 - 0.05$ eV) and develop 
a technology to go towards the normal-hierarchy mass scale 
($\langle m_{\nu}\rangle \sim 0.01$ eV). The experimental sensitivity 
requested to explore the inverted-hierarchy region (for the 
nuclei with the highest decay probability) is on the level of 
$\lim T_{1/2}\sim10^{26}-10^{27}$~y. The sensitivity requirements 
are even much stronger taking into account certain problems of 
nuclear-matrix-element calculation accuracy and a possible 
quenching of the axial vector coupling constant ($g_A$)~\cite{Barea:2012}. 

In light of the foregoing, cryogenic scintillating bolometers look 
very promising detectors for the next generation $0\nu2\beta$ 
experiments thanks to their high energy resolution (a few keV), $80-90$~\% detection efficiency, and excellent particle 
identification ability \cite{Pirro:2006,Giuliani:2012b,Artusa:2014}. 
The isotope $^{100}$Mo is one of the most promising $2\beta$ nuclei taking 
into account the high energy of the decay 
($Q_{2\beta}=3034.40(17)$ keV \cite{Rahaman:2008}), the 
comparatively high natural isotopic abundance ($\delta=9.744(65)\%$ 
\cite{Meija:2016}), and the possibility of isotopical 
separation by centrifugation in a large amount. The recent 
calculations of nuclear matrix elements for the $0\nu2\beta$ decay 
of $^{100}$Mo give comparatively ``short'' half-life in the range of 
$T^{0\nu2\beta}_{1/2}\approx(0.7-1.7)\times 10^{26}$~y 
\cite{Rodriguez:2010,Simkovic:2013,Hyvarinen:2015,Barea:2015} (for 
an effective Majorana neutrino mass equal to 0.05 eV, assuming the standard value of 
the axial vector coupling constant $g_{A}=1.27$, and using the 
recent calculations of the phase-space factor from Ref.~\cite{Kotila:2012}). 

The availability of molybdenum-containing scintillators to be operated 
as cryogenic scintillating bolometers is an important practical 
advantage of $^{100}$Mo. Recently, lithium molybdate 
(Li$_2$MoO$_4$) crystal scintillators were successfully tested as scintillating 
bolometers \cite{Barinova:2010,Cardani:2013}. Subsequently, a technique to 
grow large-volume, high-quality Li$_2$MoO$_4$ crystal 
scintillators with low radioactive contamination -- embedding also enriched $^{100}$Mo -- 
was developed in the framework of the LUMINEU~\cite{Tenconi:2015,Weblumineu} 
and ISOTTA~\cite{Webisotta} projects with outstanding results~\cite{Bekker:2016,Preprint:2016}. 
This makes this material very promising for $0\nu2\beta$ experiments with $^{100}$Mo. 

However, random coinciding events, especially of the $2\nu2\beta$ 
decay of $^{100}$Mo, can produce background due to the poor time 
resolution of cryogenic detectors 
\cite{Beeman:2012,Chernyak:2012}. This effect can be a major source 
of background in the region of interest on the level of 
$\sim 10^{-3}$ counts/(keV$\cdot$kg$\cdot$y) \cite{Chernyak:2014}, 
depending on the detector volume and performance, and on the 
data-analysis approach. As it was demonstrated in \cite{Chernyak:2014} the 
rejection efficiency substantially depends on the time properties 
and signal-to-noise ratio of the detector. Here we analyze 
the advantages of a cryogenic light detector operated with Neganov-Luke 
amplification \cite{Neganov:1981,Luke:1988} to reject pile-up 
signals in a Li$_2$MoO$_4$-based scintillating bolometer. 

\section{Neganov-Luke light detectors}

In scintillating bolometers for $0\nu2\beta$ search, employed in 
the LUMINEU \cite{Tenconi:2015} and LUCIFER \cite{Beeman:2013a} 
projects, the light emitted by the scintillating crystal is 
detected and measured by an auxiliary bolometer, consisting of a 
pure Ge wafer, working as light collector and energy absorber. The 
wafer is coupled to a neutron transmutation doped (NTD) Ge 
thermistor, serving as a temperature sensor. Details on these 
light detectors can be found in Refs. 
\cite{Tenconi:2012,Beeman:2013b}. In this work, we will consider 
Ge wafers with a diameter of 44 mm and a thickness of 0.17 mm, as 
those used by the LUCIFER and LUMINEU collaborations in their 
pilot experiments. In general, the performances of these light 
detectors present a significant spread, due to a difficult 
reproduction of the thermal couplings among detector elements, but 
they are always largely sufficient to separate $\alpha$ and 
$\beta$ particles in the region of interest for 
$0\nu2\beta$ decay of $^{100}$Mo (around 3034 keV), exploiting 
their different light yields~\cite{Bekker:2016,Beeman:2012}.

An average-performance detector based on the NTD Ge technology in the LUMINEU or LUCIFER context, in optimized 
noise conditions, has typically a baseline width of the order of 
$\sim 100$~eV rms, even if occasionally much better values -- 
around $40-50$~eV rms -- are observed 
\cite{Tenconi:2012,Beeman:2013b}. We will take conservatively the 
former value for the discussion that will follow. The light 
collected in a Li$_2$MoO$_4$ scintillating bolometer corresponds 
to an energy deposition in the light detector of about 1~keV when 
1~MeV total energy is released by electrons in the scintillator. In previous 
tests with this compound, lower values were observed 
(of the order of 0.4 keV/MeV \cite{Cardani:2013} or 0.7 keV/MeV 
\cite{Bekker:2016}), but recently it was possible to obtain 
systematically light yields close to $\sim 1$~keV/MeV thanks to an improved crystal 
quality~\cite{Preprint:2016}. The signal in the light detector induced by a 
$0\nu2\beta$ event corresponds therefore to about 3~keV energy. 
Consequently, a typical signal-to-noise ratio (defined as the ratio of 
the signal amplitude to the standard deviation of the 
noise baseline) of light detectors operated with Li$_2$MoO$_4$ 
crystal scintillators is $\sim 30$ \cite{Chernyak:2014}. 

The signal-to-noise ratio in the light channel can be enhanced by a large factor by 
exploiting the Neganov-Luke effect \cite{Neganov:1981,Luke:1988}, 
keeping essentially the same light-detector structure and 
materials and especially the same temperature sensor. The latter 
point is of great importance in view of a large scale experiment 
like that proposed by the CUPID group of interest~\cite{cupid}, 
since the NTD Ge readout is very simple and involves only 
well-established room-temperature electronics with easy 
channel multiplication \cite{Arnaboldi:2006}. 

\begin{figure}
\centering
\includegraphics[width=0.45\textwidth]{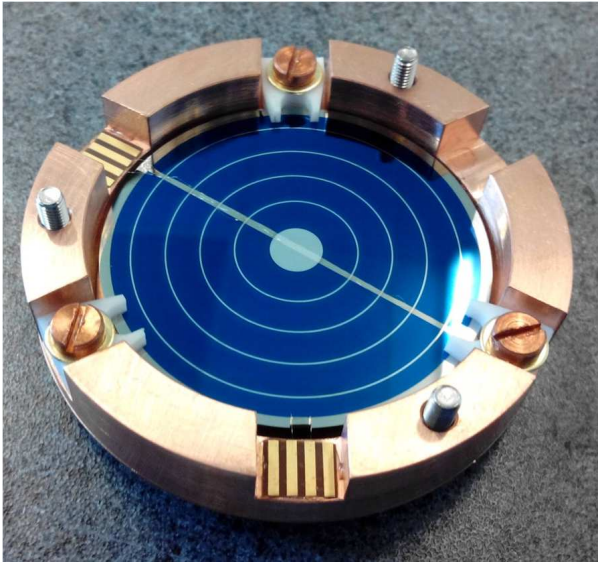}
\caption{A Neganov-Luke light detector fabricated at CSNSM. 
It consists of a 44-mm-diameter Ge disk provided with a set of 
concentric Al electrodes with a pitch of about 3.7 mm and coated with a 70-nm-thick SiO antireflective film (blue color area). The NTD Ge thermistor 
for the temperature readout is visible in the lower part of the photograph 
as a small chip attached at the Ge disk close to the edge. The visible uncoated diametrical band allows for contacting the annular electrodes.}
\label{Fig:photo-light-detector}
\end{figure}

The Neganov-Luke effect consists in a heat-mediated 
voltage-assisted measurement of the charge developed in a 
semiconductor detector by impinging radiation. It enables the 
detection of very small amount of charges, down to a few 
electron-hole pairs, with much better sensitivities with respect 
to the conventional readout based on charge-sensitive amplifiers. 
To this aim, the semiconductor bolometric absorber is provided 
with electrodes on its surface, which are used to apply an 
electric field in the absorber volume. The work done by the fields 
on the drifting charges can be detected in form of heat by NTD Ge thermistors. In the case of our 
44-mm-diameter Ge disk, the electrodes are a set of concentric Al 
rings deposited on one surface by evaporation with a shadow mask. 
The rings are electrically connected by means of ultrasonic wedge 
bonding with an alternate pattern. This allows applying a given 
voltage drop between any couple of adjacent rings and producing an 
electric field parallel to the surface. A photograph of the device 
is shown in Fig.~\ref{Fig:photo-light-detector}. The use of a ring 
structure, instead of a peripheral and central electrode only, 
enables to increase the charge-collecting electric field for a 
given applied voltage and to decrease the path length of the 
charges towards the electrodes, implying a lower trapping probability.

Several Neganov-Luke light detectors, with different inter-electrode 
pitches and adding in some samples a SiO antireflective coating 
with a thickness of 70 nm \cite{Mancuso:2014}, were fabricated, tested and characterized. One of these 
devices has been used succesfully to detect the tiny Cherenkov 
light emitted by a TeO$_2$ crystal \cite{Pattavina:2015}, in the 
framework of a R\&D activity for an improvement of the 
$0\nu2\beta$ detectors used in the CUORE experiment 
\cite{Artusa:2015} and designed to be used in its proposed 
evolution CUPID \cite{cupid}. In general, we have shown that it is 
possible to apply safely $\sim 50$~V to the electrode structure 
discussed above and shown in Fig.~\ref{Fig:photo-light-detector}, 
without the development of leakage currents. Often, it is possible 
to reach $\sim 100$~V. The signal amplification achievable on LED 
pulses in the infrared range is spectacular, of the order of 
$\sim 30$. The baseline noise remains almost constant under signal Neganov-Luke 
amplification, with a slight increase observed sometimes in the 
high-voltage range ($50-100$~V). 

Of course, it is important to understand and control the noise sources which contribute to the baseline fluctuations. These are mainly due to parasitic effects, like vibrations (which induce temperature fluctuations of Ge wafer), microphonic noise (generated by the readout-wire mechanical oscillations), and, to a minor extent, intrinsic noise of the thermistor (johnson and 1/f noise) and of the preamplifier. Our rejection method could take advantage of a mitigation of these contributions, which however are not amplified by the Neganov-Luke effect, as discussed above. A more dangerous noise source that we have observed is related to the aforementioned leakage currents developed at high voltages. The associated fluctuations can contribute significantly to the noise and the Neganov-Luke effect amplifies them along with the signal. This sets an intrinsic upper limit to the achievable gain in terms of signal-to-noise ratio.  

\begin{figure}
\centering
\includegraphics[width=0.45\textwidth]{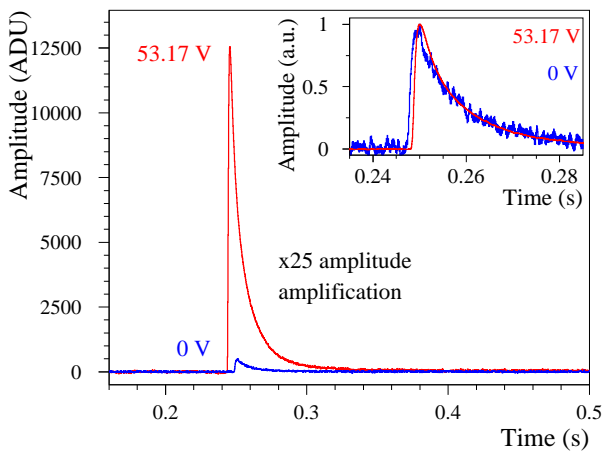}
\caption{Comparison of light-detector pulses (in ADC unit) induced 
by the asborption of a flash of light emitted by a LED 
(wavelength 1.45 $\mu$m) with (53.17 V label, alluding to the 
voltage value applied between two adjacent Al rings) and without 
(0 V label) Neganov-Luke effect. In the inset: the two pulses are 
normalized in amplitude in order to emphasize the factor $\sim 20$ 
improvement of the signal-to-noise ratio.}
\label{Fig:improvement-SNR}
\end{figure}

The best results obtained up to 
now is an improvement of the signal-to-noise ratio of $\sim 20$ 
with respect to the performance in absence of Neganov-Luke effect, 
as shown in Fig.~\ref{Fig:improvement-SNR}. We are confident 
however that this figure can be largely improved, as these results 
are very preliminary and an extensive optimization work is still to be 
done in terms of electrode configuration, deposition procedure and noise control. 
We will assume in the following that a gain of $25$ can be 
obtained, close to what already achieved and rather conservative 
with respect to the potential of this technology.

We would like to stress that the improvement in terms of signal-to-noise ratio obtained by the Neganov-Luke effect cannot increase the light-detector energy resolution beyond the limit set by the photon statistics. At the $^{100}$Mo $0\nu2\beta$ characteristic energy, about 1450 scintillation photons are collected by the light detector, as $\sim 1$~keV is the total energy contained in the corresponding scintillation pulse. (The light emission from Li$_2$MoO$_4$ has an intensity maximum at $\sim 600$~nm~\cite{Bekker:2016}, which corresponds to a photon energy of 2.07 eV.) A poissonian standard deviation of $\sim$~40 photons is therefore expected, corresponding to an intrinsic limit on the energy resolution of $\sigma \approx 80$~eV. This however plays no role in our following discussion about pile-up rejection, which requires as much as possible noise-free pulses and is not affected by an energy-resolution loss due to the granularity of the energy carriers.

\section{Generation of randomly coinciding events}

In order to discriminate random coincident events in scintillating bolometers, 
it is possible to exploit pulse-shape analysis either in the heat or in the 
light channel signals. The formers are slower but feature a much better 
signal-to-noise ratio. Even if light signals can provide a significant 
discrimination with the state-of-the-art light detectors~\cite{Chernyak:2012}, 
heat signals are superior in terms of rejection efficiency~\cite{Chernyak:2014} 
as their larger signal-to-noise ratio (typically of the order of $\sim 10^3$) prevails. 
The rationale of using Neganov-Luke light detectors is to exploit the ten-time faster 
light signals in terms of rise-time with a signal-to-noise ratio 
that approaches that of the heat channel.

We have investigated the rejection efficiency that can be obtained with this 
light-detector technology, and consequently the final background rate in 
the region of interest due to random coincidences of two-neutrino $2\beta$ decay, 
assuming a single module consisting of a cylindrical Li$_2$$^{100}$MoO$_4$ crystal 
with a diameter of 44~mm and a height of 60~mm, coupled to a Neganov-Luke light detector 
like the one described in the previous section (an array of single modules with 
these features will be tested in the framework of the CUPID R\&D program). 
Assuming 100\% enrichment, such a crystal contains $9.4 \times 10 ^{23}$ $^{100}$Mo nuclei. 
The consequent random-coincidence background rate $B_{rc}$ amounts to~\cite{Chernyak:2012}:

\begin{equation}
B_{rc} \left[ \rm{counts/(keV} \cdot \rm{ kg} \cdot \rm{y)} \right] 
\approx 3.37 \times 10^{-4} \cdot \left[ T_R/\rm{1 \, ms} \right], 
\label{eq:bkg}
\end{equation} 

\noindent
where $T_R$ is the pulse-pair resolving time. The assumption underlying 
this formula is that two signals separated by an interval shorter than $T_R$ 
will be analysed as a single pulse with an amplitude given by the sum of 
the two individual ones, while they will be recognized as double if 
the time separation is longer than $T_R$.

Ten thousands of noise baseline samples and a scintillation reference pulse were used to generate sets of single and randomly coincident pulses. 
The noise samples were acquired by a real Neganov-Luke light-detector baseline with a sampling frequency of 20~kSPS. In order to build the scintillation reference pulse, we took 40 individual scintillation pulses (sampled with 1~kSPS) from an ordinary light detector based on a Ge disk instrumented with an NTD Ge thermistor, in a setup similar to that described in Ref.~\cite{Bekker:2016}. We stress that this device is identical to those equipped with Al rings to exploit the Neganov-Luke effect. The light detector was coupled to a 240~g Li$_2$MoO$_4$ scintillating bolometer. The scintillation pulses used to build the reference pulse corresponded to $\gamma$ and $\beta$ events with energies in the 1.5-2.6 MeV range in the Li$_2$MoO$_4$ scintillator. The reference pulse was obtained by fitting the average pulse built on these 40 individual light signals and therefore it represents faithfully the shape of a scintillation signal. The phenomenological fitting function is a sum of three exponentials with 3 free amplitudes and 3 free time constants. It is not based on a detector-response model, but it represents very accurately the pulse shape. The rise-time of the reference pulse (defined as the time to change the pulse amplitude on the front edge from 10\% to 90\% of its maximum) is $\tau_{\rm rise} \approx3$~ms, while the decay time (the time to change the pulse amplitude on the pulse decay from 90\% to 30\% of its maximum) is $\tau_{\rm decay} \approx14$ ms. 

To generate randomly coinciding light signals corresponding to overlapping 
heat signals (assuming 1 keV/MeV light-to-heat ratio as discussed above) in 
the region of the $^{100}$Mo $Q_{2\beta}$ value, the amplitude of 
the first pulse $A_1$ was obtained from the $2\nu2\beta$ 
distribution of $^{100}$Mo, while the amplitude of the second 
pulse was chosen so that the total pulse energy was $Q_{2\beta} 
+ \Delta E$, where $\Delta E$ is a random component in the energy 
interval $[-5, +5]$ keV~\cite{Chernyak:2012}. Ten thousands of single pulses 
and ten thousands of coinciding signals were randomly generated in the 
time interval from 0 to $3.3\cdot \tau_{\rm rise}$. The choice of the factor 3.3 is arbitrary. It garantees that pulses separated by a longer interval are recognized as double with 100\% efficiency, as discussed below. As far as this condition is respected, the final results on the rejection efficiency do not depend on the value of this factor.

\begin{figure}
\centering
\includegraphics[width=0.5\textwidth]{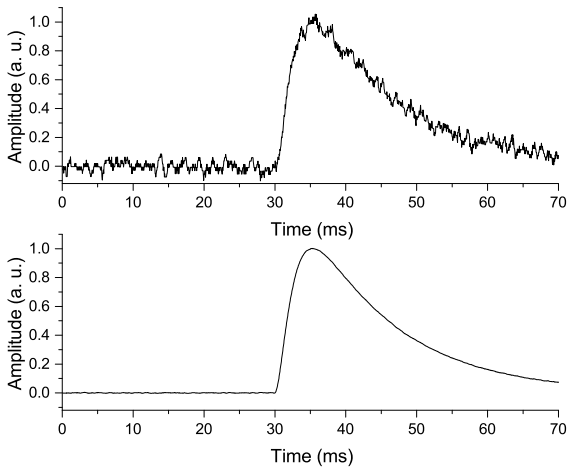}
\caption{Examples of generated light pulses with signal-to-noise 
ratio 30 (upper panel) and 750 (lower panel).}
\label{fig-pls}
\end{figure}

An example of light signal obtained with an ordinary cryogenic light detector, 
with a signal-to-noise ratio around 30 as discussed above, is given in 
the upper panel of Fig.~\ref{fig-pls}. As discussed in the previous section, 
we assume that a light detector based on Neganov-Luke effect can improve 
this value up to $25$ times, leading to a signal-to-noise ratio of $750$. 
A single pulse in these conditions is also shown in Fig.~\ref{fig-pls} (lower panel). 

\section{Results and discussion}

The mean-time method was applied to discriminate randomly 
coincident events. The mean-time parameter $\langle t \rangle$ was 
calculated for each pulse $f(t_k)$ by using the following formula: 

$$\langle t \rangle = \sum {f(t_k)\cdot t_k}/\sum {f(t_k)},$$

\noindent where the sum is over time channels $k$, starting from 
the start of a pulse and up to a certain time. 

\begin{figure}
\centering
\includegraphics[width=0.55\textwidth]{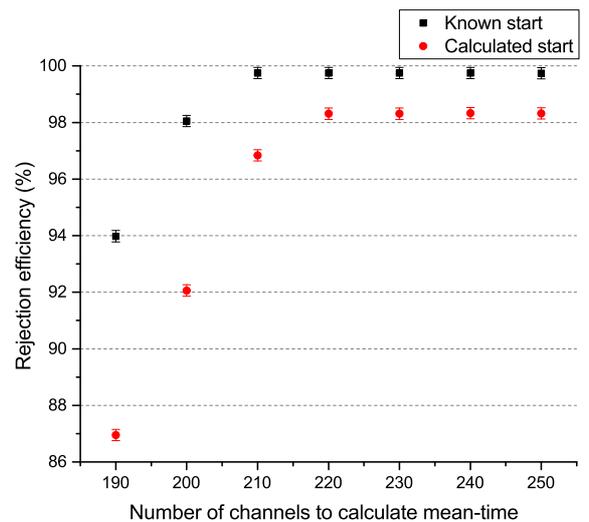}
\caption{Dependence of the rejection efficiency of the mean-time 
method on the number of channels to calculate the mean-time 
parameter $\langle t \rangle$. The analysis was performed for the 
Li$_2$MoO$_4$ light signals with 3 ms rise-time and the 
signal-to-noise ratio 750 under two conditions of the signal-origin determination: (squares) start positions of the signals 
known from the generation procedure; (circles) start positions 
found by the pulse profile analysis. One channel is 0.05 ms.}
\label{chan-n}
\end{figure}

The number of channels used to calculate the mean-time parameter 
was optimized to achieve an as-high-as-possible rejection efficiency, defined as the number of rejected coinciding 
events divided by the number of randomly generated events in the time interval 
$3.3\cdot \tau_{\rm rise} \sim 10$ ms in the light channel. We have verified that 
when two pulses are separated by an interval longer than $\sim 10$~ms 
(corresponding to about three times the light-signal rise-time) 
they are recognized as double with 100\% efficiency. An example of 
the mean-time method optimization is presented in Fig.~\ref{chan-n}. 
The rejection efficiency reaches its maximum when the mean-time parameter 
is calculated from the signal origin to approximately $220-250$ channels 
corresponding roughly to $\sim \tau_{\rm decay}$.

\begin{figure}
\centering
\includegraphics[width=0.55\textwidth]{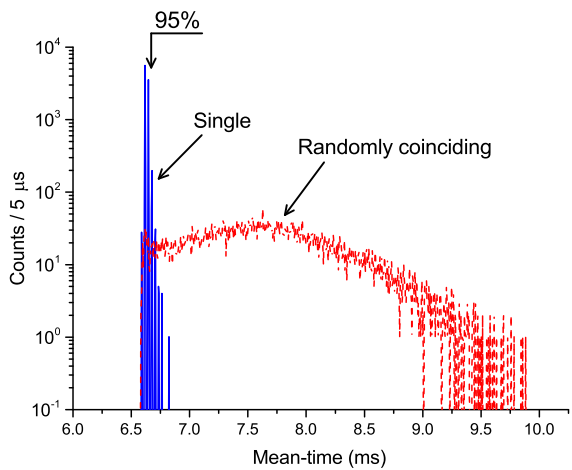}
\caption{Distribution of the mean-time parameter for single and 
randomly coinciding light pulses with a rise-time $\tau_{\rm rise} = 3$~ms and 
signal-to-noise ratio 750. The rejection efficiency of randomly 
coinciding pulses, separated by time intervals equally distributed 
in the range [$0 - 3.3\cdot \tau_{\rm rise}$], is 98.3\% under the requirement to accept 95\% of single events. The signal origin is determined by the pulse-profile analysis.}
\label{MT-distrib}
\end{figure}

The distributions of the mean-time parameters for single and 
pile-up events, generated for a light detector with a signal-to-noise 
ratio of 750, are presented in Fig. \ref{MT-distrib}. The 
rejection efficiency of randomly coinciding pulses, under the 
requirement to detect 95\% of single events, is 98.3\%. 

The rejection efficiencies computed by the simulations are presented in Table~\ref{tab1}, 
where the results obtained with an ordinary light detector are also given for comparison.

\begin{table}[!htb]
\caption{Rejection efficiency of randomly coinciding $2\nu2\beta$ 
events achievable by pulse-shape discrimination using light signals for 
different signal-to-noise ratios (without and with Neganov-Luke effect) 
and two conditions of the signal origin determination: start of the signals 
known from the generation procedure and start position found by the pulse-profile analysis.}
\centering
\begin{tabular}{|l|l|l|l|l|}
 \hline

 Channel,   & Signal-to-noise & Start       & Rejection  \\
 rise-time  & ratio          & position    & efficiency (\%)\\

 \hline

 \multirow{4}{*}{Light, 3 ms} & \multirow{2}{*}{30} & Known & $98.4\pm0.2$ \\
 \cline{3-4}
 ~                                                & & Found & $86.3\pm0.2$ \\
 \cline{2-3}
 \cline{3-4}
 ~                                                  & \multirow{2}{*}{750} & Known & $99.8\pm0.2$ \\
 \cline{3-4}
 ~                                                 & & Found & $98.3\pm0.2$ \\
 \hline

\end{tabular}
\label{tab1}
\end{table}

\begin{figure}
\centering
\includegraphics[width=0.55\textwidth]{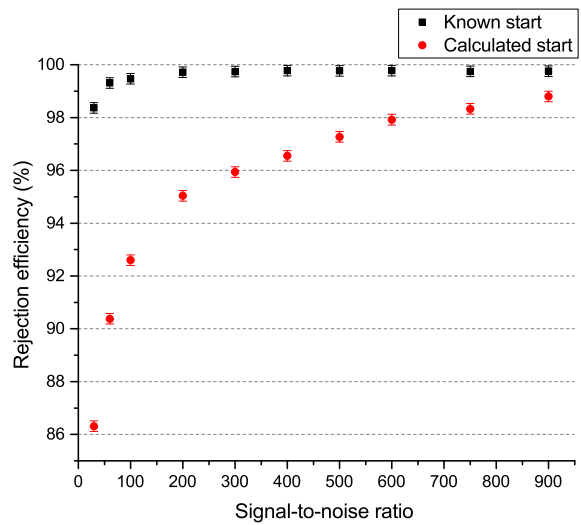}
\caption{Dependence of the rejection efficiency on the 
signal-to-noise ratio obtained using the mean-time method for Li$_2$MoO$_4$ light signals under two conditions of the signal-origin
determination: (squares) start positions of the signals 
known from the generation procedure; (circles) start positions 
found by the pulse profile analysis.}
\label{signal-noise}
\end{figure}

The dependence of the rejection efficiency on the signal-to-noise 
ratio obtained using the mean-time method for the Li$_2$MoO$_4$ 
light signals was studied for start positions of the signals found by our 
algorithm (as in a real experiment), and using the exact signal start 
positions known from the generation procedure (to estimate the maximum achievable efficiency). The results are shown in Fig.~\ref{signal-noise}. The rejection efficiency depends remarkably on the accuracy of the pulse-origin 
determination, which is substantially improved by the high signal-to-noise 
ratio provided by the Neganov-Luke light-detection technology. 

In order to translate the rejection efficiencies reported in Table \ref{tab1} 
into background rate levels, we use Equation~(\ref{eq:bkg}) with $T_R = 10$~ms 
and multiply the resulting value by the complement to 1 of the rejection 
efficiencies reported in Table~\ref{tab1} for the reconstructed pulse-origin case. 
We obtain $4.6 \times 10^{-4}$ counts/(keV$\cdot$kg$\cdot$y) and 
$5.6 \times 10^{-5}$ counts/(keV$\cdot$kg$\cdot$y) for an ordinary and 
a Neganov-Luke effect light detector, respectively. It is interesting 
to compare these values with $1.1 \times 10^{-4}$ counts/(keV$\cdot$kg$\cdot$y), which is the background rate estimated for a similar-volume Zn$^{100}$MoO$_4$ 
scintillating bolometer and using the heat channel to perform pulse-shape 
discrimination~\cite{Chernyak:2014}. It is clear that with an ordinary light detector the heat signals provide a better discrimination. The situation changes drastically in favour of the use of the light signals if a Neganov-Luke amplifying technology can be implemented.

We remark that, in order to obtain the background rate achievable with our rejection method, we have to insert $T_R \approx 0.17$~ms in Equation~(\ref{eq:bkg}). In other terms, the combination of a Neganov-Luke light detector with the mean-time pulse-shape analysis technique allows achieving an effective pulse-pair resolving time of the order of only $\sim 0.17$~ms in large-volume Li$_2$MoO$_4$ scintillating bolometers.

\section{Conclusions}

Background caused by pile-up events in Li$_2$MoO$_4$ cryogenic 
scintillating bolometers, in particular by random coincidences of 
the two-neutrino $2\beta$ events of $^{100}$Mo, can be 
effectively suppressed by pulse-shape discrimination of signals 
from light detectors based on the Neganov-Luke effect. An 
advantage of the Neganov-Luke light detectors coupled to 
Li$_2$MoO$_4$ crystal scintillators is a high signal-to-noise 
ratio, up to a level of 750, as assumed in this paper on the basis 
of experimental results on prototypes of Neganov-Luke light detectors. 
The application of the mean-time
pulse-shape discrimination reduces the random-coincidence background 
down to $\sim 5.6\times 10^{-5}$ counts/(keV$\cdot$kg$\cdot$y), 
with a remarkable pile-up rejection efficiency of 98.3\% in a $0-10$~ms 
time interval, which corresponds to the typical resolving time for a light signal. 
A high signal-to-noise ratio is a crucial characteristic for a cryogenic 
light detector in order to achieve a high discrimination efficiency of 
pile-up events, mostly because this increases the accuracy of the pulse-origin 
determination, on which the efficiency depends substantially.  

In a heat-energy window of 5 keV, in agreement with the energy resolution 
of the bolometric technique, we expect therefore a background contribution from random coincidences of 
two-neutrino $2\beta$ events inferior to 1~counts/(ton$\cdot$y). As extensively discussed in the context of the 
CUPID project and in general of next-generation $0\nu2\beta$ decay experiments~\cite{Artusa:2014,Beeman:2012,cupid},  the dominant background in $^{100}$Mo-based detectors is in fact due to $2\nu2\beta$ decay, with reasonable assumptions on all the other background sources (material radiopurity and gamma, neutron and muon external radiation). Therefore, our work addresses the most critical aspect of the Li$_2$MoO$_4$ technology in terms of background and shows that it is compatible with a full exploration of the 
inverted-hierarchy region of the neutrino mass pattern if implemented in a large-scale next-generation experiment.
 
We remark that the temperature and the electronic readouts for the Neganov-Luke 
light detectors here described are identical to those used presently in CUORE, 
making the present approach particularly attractive for CUPID. We stress also that, 
in case of an array of hundreds of bolometers as that proposed in CUPID, 
the voltage to be applied to the light-detector electrodes can be delivered 
with only a pair of wires from room temperature to the cryogenic experimental space, 
since the electrode pairs of all the light detectors can be connected 
in parallel at the array level.

\begin{acknowledgements}

D.M.C., F.A.D. and V.I.T. were supported in part by the IDEATE 
International Associated Laboratory (LIA) and by project 4-2015 of 
the National Academy of Sciences of Ukraine. F.A.D. acknowledges 
the support of the Universit\'{e} Paris-Sud as ``missionaire scientifique invit\'e'' (2016). D.V.P. was partially 
supported by the P2IO LabEx (ANR-10-LABX-0038) in the framework 
``Investissements d'Avenir'' (ANR-11-IDEX-0003-01) managed by 
the Agence Nationale de la Recherche (ANR, France). 
The Neganov-Luke light detectors were developed with the partial support 
of LUMINEU, a project receiving funds from the ANR.

\end{acknowledgements}



\begin{thebibliography}{99}


 \bibitem{Saakyan:2013} R.~Saakyan, 
Annu. Rev. Nucl. Part. Sci. \textbf{63}, 503 (2013)

 \bibitem{Barabash:2015} A.S.~Barabash, 
Nucl. Phys. A \textbf{935}, 52 (2015)

 \bibitem{Vergados:2012} J.D.~Vergados, H.~Ejiri, F.~$\mathrm{\check{S}}$imkovic, 
Rep. Prog. Phys. \textbf{75}, 106301 (2012)

 \bibitem{Barea:2012} J.~Barea, J.~Kotila, F.~Iachello, 
Phys. Rev. Lett. \textbf{109}, 042501 (2012)

 \bibitem{Rodejohann:2012} W.~Rodejohann, 
J. Phys. G \textbf{39}, 124008 (2012)

 \bibitem{Deppisch:2012} F.F.~Deppisch, M.~Hirsch, H. P\"{a}s, 
J. Phys. G \textbf{39}, 124007 (2012)

 \bibitem{Bilenky:2015}  S.M.~Bilenky, C.~Giunti, 
Int. J. Mod. Phys. A \textbf{30}, 1530001 (2015)

 \bibitem{Pas:2015} H.~P\"{a}s and W.~Rodejohann, 
New J. Phys. \textbf{17}, 115010 (2015)

 \bibitem{Elliott:2012} S.R.~Elliott, 
Mod. Phys. Lett. A \textbf{27}, 123009 (2012)

 \bibitem{Giuliani:2012a} A.~Giuliani, A.~Poves, 
AHEP \textbf{2012}, 857016 (2012)

 \bibitem{Cremonesi:2014} O.~Cremonesi, M.~Pavan, 
AHEP \textbf{2014}, 951432 (2014)

 \bibitem{Sarazin:2015} X.~Sarazin, 
J. Phys.: Conf. Ser. \textbf{593}, 012006 (2015)

 \bibitem{Delloro:2016} S.~Dell'Oro, S. Marcocci, M. Viel, F. Vissani, 
AHEP \textbf{2016}, 2162659 (2016)

 \bibitem{Kamland:2016} A.~Gando et al., 
Phys. Rev. Lett. \textbf{117}, 082503 (2016)

\bibitem{Pirro:2006} S.~Pirro et al., 
Phys. At. Nucl. \textbf{69}, 2109 (2006)

\bibitem{Giuliani:2012b} A.~Giuliani, 
J. Low Temp. Phys. \textbf{167}, 991 (2012)

\bibitem{Artusa:2014} D.R.~Artusa et al., 
Eur. Phys. J. C \textbf{74}, 3096 (2014)

\bibitem{Rahaman:2008} S.~Rahaman et al., 
Phys. Lett. B \textbf{662}, 111 (2008)

\bibitem{Meija:2016} J. Meija et al., Pure Appl. Chem. 88, 293 (2016)

\bibitem{Rodriguez:2010} T.R.~Rodriguez, G.~Martinez-Pinedo, 
Phys. Rev. Lett. \textbf{105}, 252503 (2010)

 \bibitem{Simkovic:2013} F.~$\textrm{\v{S}}$imkovic et al., 
Phys. Rev. C \textbf{87}, 045501 (2013)

 \bibitem{Hyvarinen:2015} J.~Hyv\"{a}rinen, J.~Suhonen, 
Phys. Rev. C \textbf{91}, 024613 (2015)

 \bibitem{Barea:2015} J.~Barea, J.~Kotila, F.~Iachello, 
Phys. Rev. C \textbf{91}, 034304 (2015)

 \bibitem{Kotila:2012} J.~Kotila, F.~Iachello, 
Phys. Rev. C \textbf{85}, 034316 (2012)

 \bibitem{Barinova:2010} O.P.~Barinova et al., 
Nucl. Instrum. Meth. A \textbf{613}, 54 (2010)

 \bibitem{Cardani:2013} L.~Cardani et al., 
JINST \textbf{8}, P10002 (2013)

 \bibitem{Tenconi:2015} M.~Tenconi et al., 
Physics Procedia \textbf{61}, 782 (2015)

 \bibitem{Weblumineu} http://lumineu.in2p3.fr/

 \bibitem{Webisotta} http://isotta.in2p3.fr/

 \bibitem{Bekker:2016} T.B.~Bekker et al., 
Astropart. Phys. \textbf{72}, 38 (2016)

 \bibitem{Preprint:2016} E.~Armengaud et al., 
Searching for neutrinoless double beta decay with Li$_2$MoO$_4$-based scintillating bolometers, 
in preparation to Phys. Lett. B (2016)

 \bibitem{Beeman:2012} J.W.~Beeman et al., 
Phys. Lett. B \textbf{710}, 318 (2012)

 \bibitem{Chernyak:2012} D.M.~Chernyak et al., 
Eur. Phys. J. C \textbf{72}, 1989 (2012)

 \bibitem{Chernyak:2014} D.M.~Chernyak et al., 
Eur. Phys. J. C \textbf{74}, 2913 (2014)

 \bibitem{Neganov:1981} B.S.~Neganov, V.N.~Trofimov, 
USSR patent No. 1037771 (1981)

 \bibitem{Luke:1988} P.N.~Luke, 
J. Appl. Phys. \textbf{64}, 6858 (1988)

 \bibitem{Beeman:2013a} J.W.~Beeman et al., 
AHEP \textbf{2013}, 237973 (2013)

 \bibitem{Tenconi:2012} M.~Tenconi et al., 
PoS (PhotoDet 2012) 072

 \bibitem{Beeman:2013b} J.W.~Beeman et al, 
JINST \textbf{8}, P07021 (2013)

 \bibitem{cupid} The CUPID Interest Group, 
arXiv:1504.03599v1[physics.ins-det]

 \bibitem{Arnaboldi:2006} C.~Arnaboldi et al., 
Nucl. Instrum. Meth. A \textbf{559}, 826 (2006)

\bibitem{Mancuso:2014} M.~Mancuso et al.,  
EPJ Web of Conferences \textbf{65}, 04003 (2014)

\bibitem{Pattavina:2015} L.~Pattavina et al., 
J. Low Temp. Phys. \textbf{184}, 286 (2016)

\bibitem{Artusa:2015} D.R.~Artusa et al., 
AHEP \textbf{2015}, 879871 (2015)


\end{thebibliography}
\end{document}